\documentclass[aps,prd,reprint,balancelastpage,nofootinbib,preprintnumbers,amsmath,amssymb,superscriptaddress,longbibliography]{revtex4-1}
\usepackage[
colorlinks=true,        
citecolor=blue,         
linkcolor=blue,         
urlcolor=blue           
]{hyperref}             
\usepackage{bm}         
\usepackage{xcolor}     
\usepackage{orcidlink}
\usepackage{units}
\usepackage{graphicx}
\usepackage{caption}
\usepackage{ragged2e}
\usepackage{multirow}
\usepackage[export]{adjustbox}
\usepackage{threeparttable}
\usepackage{makecell}
\newcommand{\nc}{\newcommand*}

\nc{\Eq}[1]{Eq.~\eqref{#1}}     
\nc{\Fig}[1]{Fig.~\ref{#1}}     
\nc{\Table}[1]{Table.~\ref{#1}}  
\nc{\Sec}[1]{Sec.~\ref{#1}}     
\def\({\left(}
\def\){\right)}
\def\[{\left[}
\def\]{\right]}
\def\e{\begin{equation}}
\def\q{\end{equation}}
\def\m{\begin{eqnarray}}
\def\n{\end{eqnarray}}

\begin{document}

\title{Beyond the Simple Power Law: A Bayesian Analysis of 897 Pulsar Spectra}


\author{Qingzheng Gao\orcidlink{0009-0004-2097-8548}}
\affiliation{Institute for Frontier in Astronomy and Astrophysics \& Faculty of Arts and Sciences, Beijing Normal University, Zhuhai 519087, China}
\affiliation{Khoury College of Computer Sciences, Northeastern University, San Jose, CA 95113, USA}

\author{Xiao-Jin Liu\orcidlink{0000-0002-2187-4087}}
\affiliation{Institute for Frontier in Astronomy and Astrophysics \& Faculty of Arts and Sciences, Beijing Normal University, Zhuhai 519087, China}

\author{Zhi-Qiang You\orcidlink{0000-0002-3309-415X}}
\email{zhiqiang.you@hnas.ac.cn}
\affiliation{Institute for Gravitational Wave Astronomy, Henan Academy of Sciences, Zhengzhou 450046, Henan, China}

\author{Zheng Li\orcidlink{0000-0002-0786-7307}}
\email{l.z@bnu.edu.cn}
\affiliation{Institute for Frontier in Astronomy and Astrophysics \& Faculty of Arts and Sciences, Beijing Normal University, Zhuhai 519087, China}

\author{Xingjiang Zhu\orcidlink{0000-0001-7049-6468}}
\email{zhuxj@bnu.edu.cn}
\affiliation{Institute for Frontier in Astronomy and Astrophysics \& Faculty of Arts and Sciences, Beijing Normal University, Zhuhai 519087, China}

\begin{abstract}

We present a comprehensive re-evaluation of pulsar radio spectra using the largest curated dataset of calibrated flux densities to date, comprising 897 pulsars, and employing a robust Bayesian framework for model comparison alongside frequentist methods. Contrary to the established consensus that pulsar spectra are predominantly simple power laws, our analysis reveals that complex spectral shapes with curvature or breaks are in fact the norm. The broken power law emerges as the most common spectral shape, accounting for 60\% of pulsars, while the simple power law describes only 13.5\%, with 68.8\% of pulsars decisively favoring curved or broken models. We further identify 74 confident gigahertz-peaked spectrum pulsars, and demonstrate that millisecond pulsars frequently exhibit spectral curvature. A key finding is that the previously reported dominance of the simple power law was largely a statistical artifact of the frequentist method used in earlier work. These findings substantially revise the prevailing view of pulsar spectra and establish a critical, model-classified foundation for future theoretical work.

\end{abstract}
\maketitle

\section{Introduction} \label{sec:intro}

Pulsars, the highly-magnetized and rapidly rotating remnants of massive stellar explosions, have been a source of fascination since their serendipitous discovery in 1967 \citep{hewish_observation_1968}.
These compact neutron stars are enveloped by magnetospheres filled with electron-positron plasma that co-rotates with the star \citep{goldreich_pulsar_1969}.
From our vantage point on Earth, they are observed as precise, periodic pulses of radio emission, akin to a cosmic lighthouse, each time their beamed radiation sweeps across our line of sight. This population of celestial clocks includes both isolated pulsars and those in binary systems \citep{kramer_double_2008}, serving as natural laboratories for studying extreme physics.
Despite decades of study, the fundamental physical mechanisms governing their emission remain elusive, though they are thought to be intimately linked to properties such as rotation, magnetic field configuration, internal structure, and the surrounding environment.

The quest to understand the pulsar emission mechanism has been pursued through a variety of methodological approaches over the past several decades. These efforts can be broadly categorized: some studies focus on in-depth analysis of remarkable individual objects to uncover specific physical phenomena \citep{karastergiou_transient_2011,desvignes_radio_2019}, while others seek to generalize findings by examining populations of pulsars that share common properties, such as age or magnetic field strength \citep{johnston_galactic_2020,lower_impact_2021}. A third, crucial approach involves large-scale statistical studies that aim to identify macroscopic trends and correlations among observable parameters across hundreds of pulsars \cite[e.g.,][]{sieber_pulsar_1973,Jankowski_2018}.

A critical avenue for understanding pulsar emission is the study of their integrated radio flux density spectra---the variation of flux density with observing frequency. Measuring the flux density not only allows for the calculation of radio luminosity and energy output but also provides key insights into the underlying emission processes. However, obtaining accurate measurements is challenging. Interstellar scintillation and instrumental systematic errors can cause flux density to fluctuate by factors of 2 up to an order of magnitude over timescales ranging from weeks to months \citep{swainston_discovery_2021}. Consequently, high-precision measurements require observations that extend beyond the scintillation timescale, coupled with rigorous error analysis and advancing calibration techniques. Thanks to such sustained efforts, flux density measurements for thousands of pulsars have now been accumulated in various catalogues and databases \cite[e.g.,][]{manchester_australia_2005,swainston_pulsar_spectra_2022}.

It has long been thought that pulsar spectra often approximately follow a power-law distribution, with a characteristic spectral index $\alpha$ ranging from $-2$ to $-1$ \cite[e.g.,][]{Lorimer_1995,xilouris_emission_1996,maron_pulsar_2000,bates_pulsar_2013,Jankowski_2018}. However, a significant number of pulsars exhibit more complex spectral shapes, including breaks or turn-overs \citep{sieber_pulsar_1973}. The standard analytical models used to describe these spectra include, beyond the simple power law, the broken power law and the log-parabolic spectrum. The latter two are particularly useful for modeling gigahertz-peaked spectra (GPS), where the flux density peaks around 1 GHz \citep{xilouris_emission_1996,bates_pulsar_2013,Dembska_2014}.

Previous population studies of pulsar flux density spectra have laid important groundwork. For instance, \citet{Jankowski_2018} performed a spectral analysis of 441 pulsars, concluding that the simple power law was dominant (79\%) based on the Akaike Information Criterion (AIC) for model comparison. More recently, \citet{Posselt_2023} presented flux measurements to 1,170 pulsars and performed power-law fits of pulsar spectra.
We build upon previous work in the following ways.
First, from an expanded dataset of 3,345 pulsars, we select 897 high-quality sources with at least four flux density measurements spanning a factor of two in frequency.
Second, we employ a Bayesian framework for model comparison, in addition to replicating the frequentist approach of \citet{Jankowski_2018}.
Third, we systematically evaluate six spectral models: the simple power law, broken power law, low-frequency turn-over power law, high-frequency cut-off power law, double turn-over spectrum, and log-parabolic spectrum.

This paper is organized as follows.
In \Sec{sec:methods}, we introduce our data-compilation procedures, statistical analysis methods and pulsar spectral models.
In \Sec{sec:results}, we present our analysis results from both Bayesian and frequentist methods.
In \Sec{sec:discuss}, we compare our results with previous works and discuss their implications.
Lastly, we present concluding remarks in \Sec{sec:conclude}.

\section{Data and Methods} \label{sec:methods}

\subsection{Data compilation}

Accurate pulsar flux density measurements are affected by several factors, including interstellar scintillation, intrinsic temporal variability, and differences in instrumental calibration across observatories. While some surveys estimate flux density using the radiometer equation with basic calibration, others employ more robust techniques that incorporate pulsed noise signals or celestial reference sources. To ensure the reliability of our spectral analysis, this study exclusively uses calibrated flux density measurements.

Our primary data source is the \textsc{pulsar\_spectra}\footnote{\url{https://github.com/NickSwainston/pulsar_spectra}} Python package (version 2.1.0) \cite{swainston_pulsar_spectra_2022}, a well-maintained and comprehensive catalog of pulsar flux densities. We manually reviewed the literature compiled within this repository and removed entries originating from uncalibrated publications. To incorporate recent measurements and expand this base dataset, we supplemented it with data from the following key studies:
\citet{sieber_pulsar_1973,maron_pulsar_2000,levin_radio-loud_2010,anderson_multi-wavelength_2012,gordon_quick_2021,Spiewak_2022,anumarlapudi_characterizing_2023,Posselt_2023}.
A complete list of all references included in our dataset is provided in the \textsc{spectral\_fit}\footnote{\url{https://github.com/QZGao/spectral_fit}} repository.

The resulting consolidated dataset represents an extensive collection of radio flux density measurements. In its raw form, it contains observations for 2,310 unique pulsars. To characterize the dataset with respect to spectral completeness, we count the number of distinct frequency measurements per pulsar: 1,667 pulsars have more than four measurements, 1,607 have more than five, and 1,537 have more than six.

For model fitting and comparison, we apply the quality criteria adopted by \citet{Jankowski_2018}. We select only those pulsars with at least four distinct frequency measurements spanning a frequency range of at least a factor of two. This result in a final high-quality sample of 897 pulsars that meet these requirements. Notably, this sample is more than twice the size of the 441 pulsars analyzed in the foundational study of \citet{Jankowski_2018}, providing a substantially larger basis for robust statistical analysis.

\subsection{Bayesian model selection} \label{sec:bayesian_model_selection}

We apply Bayesian inference to fit the flux density spectra of each pulsar in our data set. Our analysis follows the methodology described in \citet{youDeterminationBirthmassFunction2025}.
To compute posterior distributions for the model parameters and the Bayesian evidence for each model, we employ the dynamic nested sampling algorithm implemented in the Python package \textsc{dynesty}.
While the evidence includes an Occam penalty for model complexity, we treat the highest-evidence model as a convenient classification label rather than a definitive choice. We assess the strength of statistical preference using Bayes factors.

\begin{figure}
    \centering
    \includegraphics[width=\linewidth]{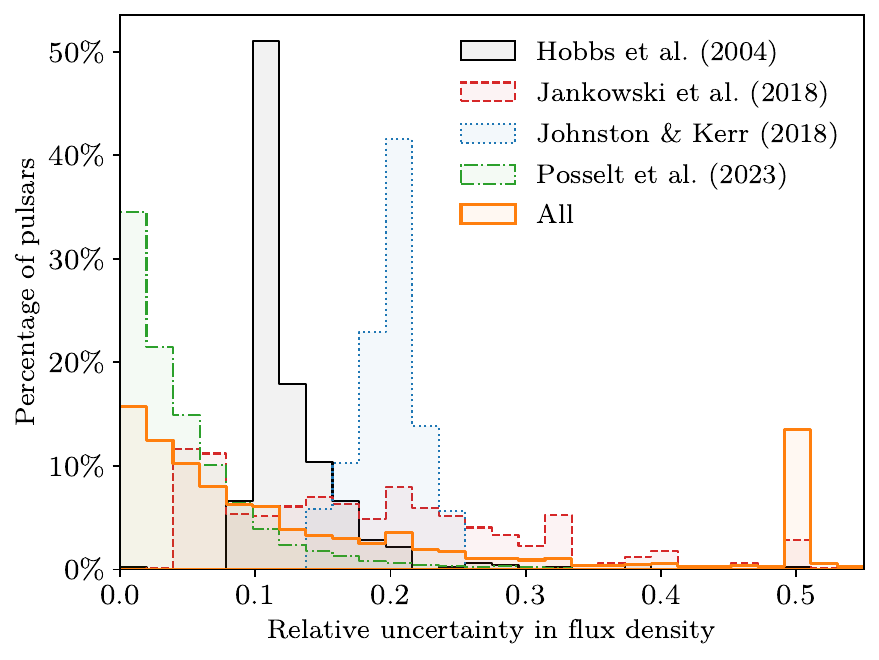}
    \caption{\justifying 
    Distributions of relative uncertainties in flux density measurements from several major references and of the full 2,310-pulsar dataset.
    In \citet{Hobbs_2004a}, 38.2\% of the measurements adopt an assumed 10\% relative uncertainty where no estimates were reported. 
    \citet{Jankowski_2018} derive uncertainties using modulation indices, producing a broad distribution. 
    \citet{Johnston_2018} do not provide uncertainties, so a 20\% relative uncertainty was assigned by \textsc{pulsar\_spectra} \citep{swainston_pulsar_spectra_2022}. 
    \citet{Posselt_2023} employ an unconventional flux density measurement method that yields relatively small uncertainties for most pulsars.
    }
    \label{fig:uncertainty_distribution}
\end{figure}

A key challenge in spectral modeling is the variability and reliability of reported flux density uncertainties. Uncertainty estimates differ widely across studies: some authors adopt fixed values---such as 50\%, 20\%, or 10\%---either alone or added in quadrature to measurement errors; others average measurements obtained over multiple epochs and use the standard deviation; and more recent work sometimes adopts the modulation index. These methodological differences reflect varying instrumentation, observational strategies, and calibration procedures, as discussed in \citet{Gitika_2023}. The absence of a unified standard is further illustrated in \Fig{fig:uncertainty_distribution}, which shows the distribution of relative uncertainties across selected studies and the full dataset.
For measurements that do not report accurate uncertainties, we adopt the assumed relative uncertainties suggested in the original papers (e.g., 10\% from \citet{Hobbs_2004a,Frail_2016}), rather than imposing new values. If no such estimates are provided, we assume a 50\% relative uncertainty, a value widely adopted in earlier works \citep{Jankowski_2018,sieber_pulsar_1973}.

In many publications, it is unclear whether the reported uncertainties adequately account for temporal variability or systematic effects. Underestimated uncertainties are particularly problematic for spectral fitting, as their quadratic contribution to the likelihood can disproportionately influence model selection.
To mitigate this, \citet{Jankowski_2018} adopted the Huber loss function to downweight outliers. However, this frequentist approach is not directly compatible with our Bayesian framework.

To account for both reported uncertainties and unmodeled systematic errors,
we instead introduce an $e_{\text{fac}}$ parameter to upscale the reported errors $\sigma_{y, i}$, following the approach used in pulsar timing array analyses. The updated Gaussian likelihood function is as follows:
\begin{equation}
    \mathcal{L} = \prod_{i}^{N} \frac{1}{\sqrt{2\pi} e_{\text{fac}} \sigma_{y, i}} \exp \left( - \frac{[f_{\theta}(x_i) - y_i]^2}{2(e_{\text{fac}} \sigma_{y, i})^2} \right),
\end{equation}
for a pulsar of $N$ measurements, where $f_{\theta}$ denotes a fitted model with parameters $\theta$. Because different studies report uncertainties in different ways, we assign an $e_{\text{fac}}$ parameter for each reference source. We assume a uniform prior for $e_{\text{fac}} \ge 1$ with an upper bound corresponding to a fixed 50\% systematic error (which is denoted as $e_{\text{quad}}$, similar to that used in pulsar timing array analysis):
\begin{equation}
    \underset{\text{per ref }k}{e_{\text{fac}}} = \sqrt{1 + \left(\frac{e_{\text{quad}}}{\min\limits_{(y_i, \sigma_{y,i}) \in k} (\sigma_{y,i}/y_{i})}\right)^2}.
\end{equation}
The upper bound on $e_{\text{quad}}$ is motivated by the distribution of relative errors shown in \Fig{fig:uncertainty_distribution}.
Specifically, 96.5\% of all reported relative uncertainties in \Fig{fig:uncertainty_distribution} are $\le 50\%$, motivating the 50\% systematic term as a conservative limit.

\subsection{Spectral models}

To characterize the diverse spectral shapes present in our pulsar sample, we evaluate six phenomenological models. These include the five models considered in \citet{Jankowski_2018} and the double turn-over model introduced by \citet{swainston_pulsar_spectra_2022}. To mitigate parameter covariance, we scale frequency with a reference value $\nu_{0}$, defined as the geometric mean of the minimum and maximum observed frequencies. This transforms the frequency variable to a dimensionless ratio $x = \nu / \nu_{0}$. Accordingly, all models below are defined in terms of $x$.

The most commonly used model is the \textit{simple power law},
\begin{equation}\label{eq:simple_power_law}
        S(x) = b x ^ \alpha,
    \end{equation}
where $\alpha$ is the spectral index.

Originally proposed to account for spectral breaks \citep{sieber_pulsar_1973}, the \textit{broken power law} model consists of two power-law segments joined at a break frequency $\nu_b$:
\begin{equation}\label{eq:broken_power_law}
    S(x) = \begin{cases}
b x^{\alpha_1}, & \textrm{if } \nu \le \nu_b, \\
b x^{\alpha_2} x_b^{\alpha_1 - \alpha_2}, & \textrm{otherwise,}
    \end{cases}
\end{equation}
where $\alpha_1$ and $\alpha_2$ are the spectral indices below and above the break, and $x_b=\nu_b/\nu_0$.

The \textit{log-parabolic spectrum}, used to identify spectral curvature and turn-overs \citep{Kuzmin_2001,kijak_spectrum_2011}, is written as
\begin{equation}\label{eq:log_parabolic_spectrum}
    S(x)=\exp\left( a x^2+b x+c \right),
\end{equation}
where $a$, $b$, and $c$ are constants.

Two additional models are derived from a general form that includes an absorption or cut-off term \citep{sieber_pulsar_1973,malofeev_mean_1980}:
\begin{equation}
    S(x)=b x^{\alpha}e^{-\tau(x)}\, .
\end{equation}

The \textit{low-frequency turn-over power law} \citep{Jankowski_2018,rajwade_gigahertz_2016} models the optical depth to produce a low-frequency turn-over, often attributed to thermal free–free absorption \citep{kijak_spectrum_2011,wilson_tools_2013}:
\begin{equation}\label{eq:low_frequency_turn_over_power_law}
        S(x)=b x^\alpha \exp\left( \frac{\alpha}{\beta} x_c^{-\beta} \right),
    \end{equation}
where $\nu_c$ is the turn-over frequency, $x_c=\nu_c/\nu_0$, and $\beta\in[0,2.1]$ controls the smoothness of the transition.

The \textit{high-frequency cut-off power law} incorporates a high-frequency cut-off by adopting an optical-depth term that produces a linear attenuation with frequency \citep{kontorovich_high-frequency_2013}. In this case, the optical depth takes the form $\tau(x) = -\ln\!\left(1 - x/x_c\right)$, leading to
\begin{equation}\label{eq:high_frequency_cut_off_power_law}
    S(x) = b x^{\alpha}\left( 1-\frac{x}{x_c} \right),
\end{equation}
where $\nu_c$ is the cut-off frequency and $x_c=\nu_c/\nu_0$. Such high-frequency behaviour is generally interpreted as arising from the pulsar's interaction with its environment rather than the intrinsic emission mechanism \citep{kijak_low-frequency_2021,ardavan_radio_2024}.

Motivated by earlier attempts to combine low- and high-frequency features \citep{sieber_pulsar_1973}, the \textit{double turn-over spectrum} \citep{swainston_pulsar_spectra_2022} includes both a low-frequency turn-over and a high-frequency cut-off:
\begin{equation}
    S(x)=b x^{\alpha} \left( 1-\frac{x}{x_{c_1}} \right) \exp\left( \frac{\alpha}{\beta} x_{c_2}^{-\beta} \right),
\end{equation}
where $\nu_{c_1}$ and $\nu_{c_2}$ are the cut-off and turn-over frequencies, respectively, with $x_{c_1}=\nu_{c_1}/\nu_0$ and $x_{c_2}=\nu_{c_2}/\nu_0$.

The prior distributions for all model parameters are listed in \Table{tab:prior_table}.

\begin{table}
    \caption{\justifying The prior distributions adopted in Bayesian inference. $U$ denotes uniform distribution, while $LU$ denotes log-uniform distribution (base 10).}
    \label{tab:prior_table}
    \centering
    \begin{tabular}{ccc}
    \hline\hline
    Model & Parameter & Prior \\
    \hline
    \multirow{2}{*}{Simple power law} & $\alpha$      & $U(-5, 0)$\\
                                      & $b$ [mJy]     & $LU(10^{-2}, 10^4)$ \\
    \hline
    \multirow{4}{*}{Broken power law} & $\nu_b$ [MHz] & $LU(\nu_{\min}, \nu_{\max})$ \\
                                      & $\alpha_1$    & $U(-5, 5)$ \\
                                      & $\alpha_2$    & $U(-5, 0)$\\
                                      & $b$ [mJy]     & $LU(10^{-2}, 10^4)$ \\
    \hline
    \multirow{3}{*}{Log-parabolic spectrum} & $a$     & $U(-5, 2)$ \\
                                      & $b$           & $U(-5, 2)$ \\
                                      & $c$           & $U(-10, 10)$ \\
    \hline
    \multirow{4}{*}{\makecell{Low-frequency turn-over\\ power law}} & $\nu_c$ [MHz] & $LU(\nu_{\min}, \nu_{\max})$ \\
                                      & $\alpha$      & $U(-5, 0)$\\
                                      & $b$ [mJy]     & $LU(10^{-2}, 10^4)$ \\
                                      & $\beta$       & $U(0, 2.1)$ \\
    \hline
    \multirow{3}{*}{\makecell{High-frequency cut-off\\ power law}} & $\nu_c$ [MHz] & $LU(\nu_{\min}, 1.5\,\nu_{\max})$ \\
                                      & $\alpha$      & $U(-5, 0)$\\
                                      & $b$ [mJy]     & $LU(10^{-2}, 10^4)$ \\
    \hline
    \multirow{5}{*}{Double turn-over spectrum} & $\nu_{c_1}$ [MHz] & $LU(\nu_{\min}, 1.5\,\nu_{\max})$ \\
                                      & $\nu_{c_2}$ [MHz] & $LU(\nu_{\min}, \nu_{\max})$ \\
                                      & $\alpha$    & $U(-5, 0)$\\
                                      & $b$ [mJy]     & $LU(10^{-2}, 10^4)$ \\
                                      & $\beta$       & $U(0, 2.1)$ \\
    \hline\hline
    \end{tabular}
\end{table}

\section{Results} \label{sec:results}

\begin{figure}
\includegraphics[width=\linewidth]{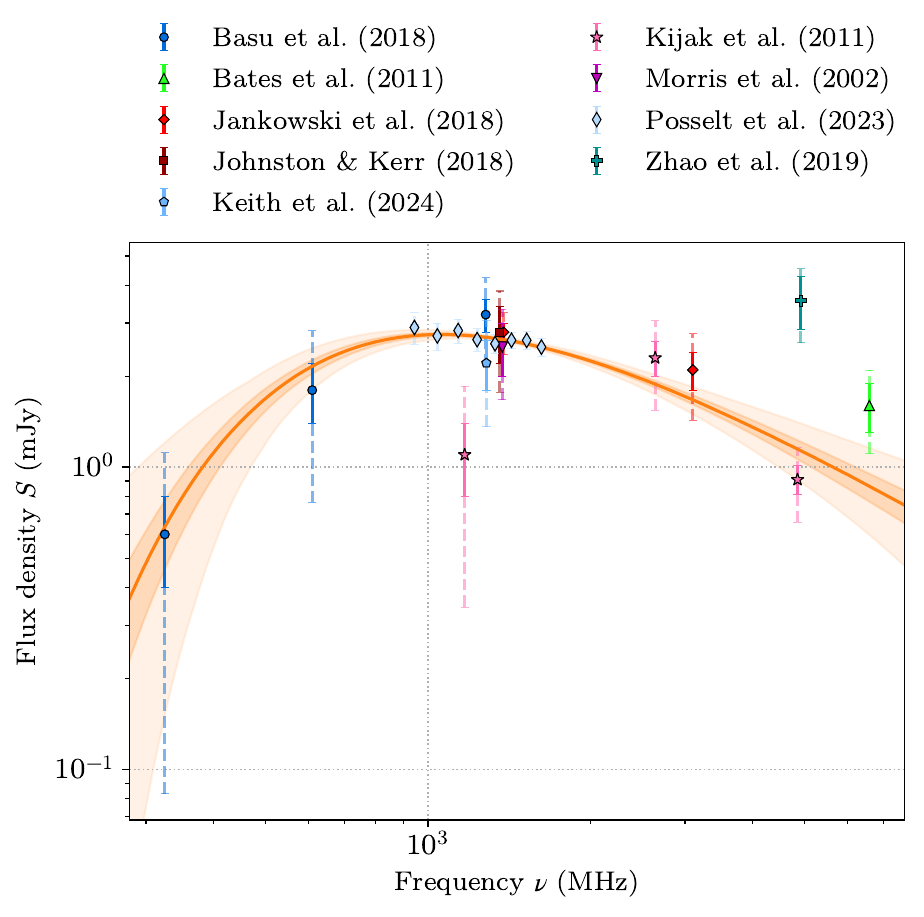}
\includegraphics[width=\linewidth]{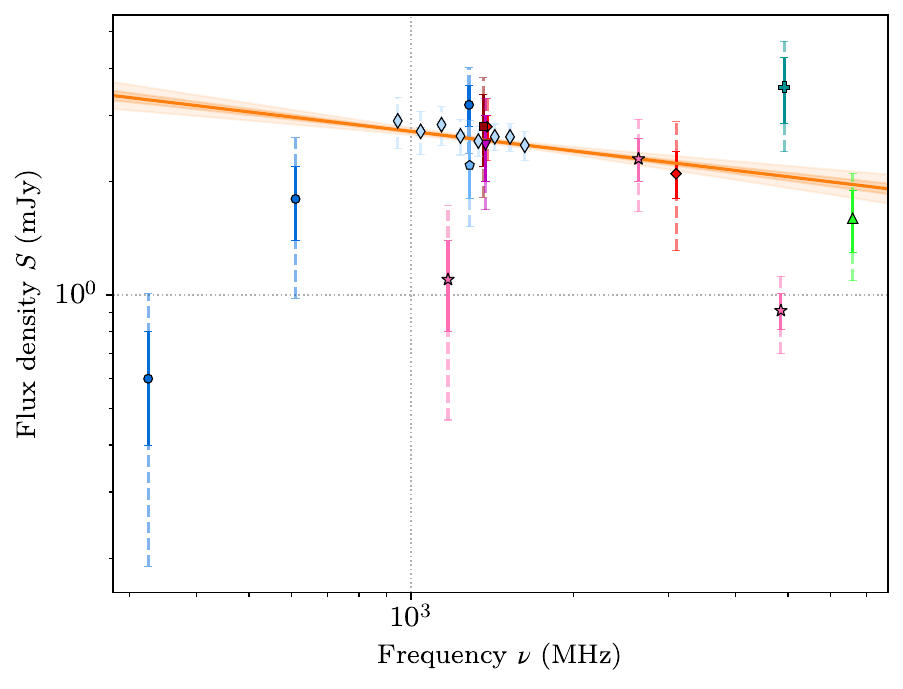}
    \caption{\justifying 
Spectrum of PSR~J1809--1917, which was previously classified as a simple power law \cite{Jankowski_2018} (lower panel), but our Bayesian analysis identifies the low-frequency turn-over power law as the best model (upper panel), with a $\ln\mathrm{BF}$ of 150 in its favor.
The solid orange line and the shaded region display the posterior predictive distribution and its associated uncertainty, with the dark and light colored region showing the 1-$\sigma$ and 3-$\sigma$ credibility interval, respectively.
Solid error bars denote the uncertainties reported in the literature, while dashed error bars represent the rescaled uncertainties with inferred $e_{\text{fac}}$ parameters. 
    }
    \label{fig:J1809-1917}
\end{figure}

\subsection{Bayesian model selection}


\begin{table*}
    \caption{\justifying Percentages of pulsars best-fit by each model among the 897 qualified pulsars. The best-fit model is defined as the one with the highest log evidence (Bayesian) or the lowest AIC (frequentist).}
    \label{tab:our_dataset_bestcat}
    \centering
    \begin{tabular}{ccc}
    \hline\hline
        Best-fit model & Highest log evidence (Bayesian) & Lowest AIC \\
\hline
Broken power law & 539 / 897 (60.1\%) & 291 / 897 (32.4\%) \\
Simple power law & 121 / 897 (13.5\%) & 242 / 897 (27.0\%) \\
Low-frequency turn-over power law & 110 / 897 (12.3\%) & 63 / 897 (7.0\%) \\
Double turn-over spectrum & 50 / 897 (5.6\%) & 30 / 897 (3.3\%) \\
High-frequency cut-off power law & 34 / 897 (3.8\%) & 86 / 897 (9.6\%) \\
Log-parabolic spectrum & 30 / 897 (3.3\%) & 171 / 897 (19.1\%) \\
Not categorized & 13 / 897 (1.4\%) & 14 / 897 (1.6\%) \\
\hline\hline
    \end{tabular}
\end{table*}

\begin{table*}
    \caption{\justifying Percentages of 884 categorized pulsars whose log Bayes factor is at least 5 relative to the second-best model or to the simple power law.}
    \label{tab:our_dataset_betterby}
    \centering
    \begin{tabular}{ccc}
    \hline\hline
        Best-fit model & \makecell{$\ln\mathrm{BF} \ge 5$ \\ (vs.\ second-best model)}& \makecell{$\ln\mathrm{BF} \ge 5$ \\ (vs.\ simple power law)}\\
        \hline
Broken power law & 355 / 884 (40.2\%) & 442 / 884 (50.0\%) \\
Low-frequency turn-over power law & 16 / 884 (1.8\%) & 65 / 884 (7.4\%) \\
Double turn-over spectrum & 35 / 884 (4.0\%) & 47 / 884 (5.3\%) \\
High-frequency cut-off power law & 18 / 884 (2.0\%) & 31 / 884 (3.5\%) \\
Log-parabolic spectrum & 12 / 884 (1.4\%) & 23 / 884 (2.6\%) \\
Simple power law & 3 / 884 (0.3\%) & --- \\
\hline\hline
    \end{tabular}
\end{table*}

We perform Bayesian model comparison by selecting the model with the highest log evidence (\Sec{sec:methods}). \Table{tab:our_dataset_bestcat} summarizes the classification results. The broken power law is the most common best-fit model (60.1\%), followed by the simple power law (13.5\%). The low-frequency turn-over, double turn-over, and high-frequency cut-off models account for 12.3\%, 5.6\%, and 3.8\% of pulsars, respectively. The log-parabolic spectrum accounts for 3.3\% of pulsars. As an illustration, Figure~\ref{fig:J1809-1917} shows the spectrum of PSR J1809--1917, for which our analysis strongly favors a low-frequency turn-over model over the simple power law previously reported.

We notice that a small fraction of pulsar spectra cannot be fitted with any of the considered models.
These pulsars are listed as ``not categorized" in \Table{tab:our_dataset_bestcat}; there are 13 pulsars belonging to this category.
Quantitatively, we classify a pulsar spectrum as ``not categorized" if more than 20\% of its measurements are outliers (at the 3-$\sigma$ or higher significance level) from the best-fit model.
We present more details on these not-categorized pulsars in Appendix \ref{sec:bad-fits}.

As noted in \Sec{sec:bayesian_model_selection}, the highest-evidence label is not necessarily a unique description. Relying solely on model evidence can be misleading when the differences are small. We therefore compute the log Bayes factor between each curved or broken model and the simple power law. A log Bayes factor of at least 5 is commonly interpreted as decisive support. As shown in \Table{tab:our_dataset_betterby}, 68.8\% of pulsars are decisively better fit ($\ln \mathrm{BF} \ge 5$) by a curved or broken model than by a simple power law, indicating that such spectra are common. Even under a more stringent threshold ($\ln \mathrm{BF} \ge 8$), 65.5\% of pulsars still favor one of the curved or broken models against the simple power law. Conversely, only 3 of the 121 pulsars that are best fit with a simple power law spectrum has a log Bayes factor over 5 against other spectral models.
This suggests that, even when the simple power law yields the highest evidence, its preference is not very strong.

We repeat the analysis using only pulsars with at least ten flux density measurements. With this larger number of data points, the spectral shape is more tightly constrained. Under this stricter requirement, 76.4\% (584 out of 764) of categorized pulsars decisively favor ($\ln \text{BF} \ge 5$) curved or broken models over the simple power law, indicating that the prevalence of such spectra is not affected by the small number of measurements.
Therefore, we conclude that curved or broken spectra are significantly more prevalent than reported by \citet{Jankowski_2018}.

\subsection{Model selection with frequentist statistics}

We also apply frequentist statistics, including the AIC with correction term, and robust regression using a Huber loss function to downweight outliers, following the approach of \citet{Jankowski_2018}. The full procedure is described in Appendix~\ref{app:reproduction}, where we also reproduce their dataset for direct comparison.

The classifications derived from AIC and robust regression are also listed in \Table{tab:our_dataset_bestcat}. Under these criteria, the broken power law remains the most common best-fit model, applying to 32.4\% of the 897 pulsars, while the simple power law accounts for 27.0\%. Thus, the broken power law is preferred in both Bayesian and frequentist analyses. Moreover, the AIC results further indicate that the simple power law is not the dominant spectral form: 71\% of pulsars are better fit by models with curvature or spectral features.
Similar to the Bayesian classification results, 1.6\% of pulsars are not categorized.

\section{Discussion}
\label{sec:discuss}
\subsection{Comparison with previous works}

Our analysis, based on a significantly expanded and thoroughly scrutinized dataset, leads to a substantive revision of the established understanding of pulsar radio spectra. The conventional wisdom, as suggested by \citet{Jankowski_2018}, held that the simple power law was the dominant spectral shape, describing approximately 79\% of pulsars. In stark contrast, we find that complex models exhibiting curvature or breaks are in fact the norm.
This paradigm shift can be attributed to two key factors: the substantial improvement in data quality and quantity, and critical methodological refinements.

\citet{Jankowski_2018} analyzed 441 pulsars, with their conclusions partly constrained by the available data at the time. Our work leverages a dataset more than twice the size (897 pulsars), incorporating a wealth of new, high-quality measurements from major surveys such as \citet{Posselt_2023} and \citet{Spiewak_2022}, which were not available previously. Crucially, our curation process excluded uncalibrated measurements, enhancing the overall reliability of the spectral fits.

A pivotal finding of our work is the identification of a methodological bias in the previous frequentist analysis. As demonstrated in Appendix \ref{app:reproduction}, the use of the corrected AIC by \citet{Jankowski_2018} inherently penalizes more complex models when the number of data points is small. For a pulsar with only 4 or 5 flux measurements, the AIC formula diverges for models with 3 or 4 parameters, effectively disqualifying them and artificially inflating the preference for the 2-parameter simple power law. When we reanalyze their dataset using the standard AIC, the prevalence of the simple power law in their own dataset drops from 64.1\% to 26.8\%, a figure that aligns far more closely with our new results. This indicates that the previously reported dominance of the simple power law was, to a significant extent, a statistical artifact.

Our conclusions are robust across independent statistical frameworks. The Bayesian model selection, which is not susceptible to the same finite-sample bias as the AIC, decisively favors curved or broken models for 68.8\% of the pulsars in our sample. Similarly, our frequentist analysis finds that 72.6\% of pulsars are better fit by non-power-law models. The consistency between these two approaches reinforces the conclusion that the simple power law is the exception rather than the rule.

We also compared our findings with the more recent work of \citet{xu_statistical_2024}, who reported that 68.51\% of pulsars follow a simple power law. This discrepancy likely stems from their use of an outdated and less comprehensive version of the \textsc{pulsar\_spectra} catalogue, which contained only 34 publications compared to the 156 in the version we employed. Our larger and more curated dataset, combined with our robust statistical treatment, provides a more complete and reliable picture of the pulsar population's spectral characteristics.

In summary, the prevailing view of pulsar spectra is updated by this work. The simplicity of the power law is not the standard case. Instead, most pulsars exhibit spectral curvature or breaks.

\subsection{Classification of millisecond pulsars}

Following the definition of \citet{levin_high_2013}, millisecond pulsars (MSPs) are those with period $P < 70$~ms and spin-down rate $\dot{P} < 10^{-17}~\mathrm{s}~\mathrm{s}^{-1}$. Periods and spin-down rates were obtained from the ATNF Pulsar Catalogue. Of the 897 pulsars analyzed in this work, 85 satisfy the MSP criteria. Their model classifications and log Bayes factor results are summarized in Tables~\ref{tab:msp_dataset_bestcat} and \ref{tab:msp_dataset_betterby}.

The Bayesian model selection reveals that for 51.2\% of MSPs (43 out of 84), the evidence decisively favors a curved or broken model over a simple power law ($\ln\mathrm{BF} \ge 5$). This finding directly contradicts the conclusion of \citet{Kuzmin_2001}, who reported no evidence for low-frequency turn-overs in their study of 30 MSPs. The discrepancy likely arises from their smaller sample size and less sensitive statistical framework, which was unable to detect the subtler curvature that our larger dataset and Bayesian methodology can resolve.

The broken power law and simple power law models are the most common spectral shapes among MSPs, fitting 42 and 20 pulsars according to the Bayesian method. These results are broadly consistent with the trends observed in the full 897-pulsar sample.

\begin{table*}
    \caption{\justifying Best-fit spectral models for 85 MSPs, based on either the highest Bayesian evidence or the lowest AIC.}
    \label{tab:msp_dataset_bestcat}
    \centering
    \begin{tabular}{ccc}
    \hline\hline
       Best-fit model & Highest log evidence (Bayesian) & Lowest AIC \\
\hline
Broken power law & 42 / 85 (49.4\%) & 22 / 85 (25.9\%) \\
Simple power law & 20 / 85 (23.5\%) & 28 / 85 (32.9\%) \\
Low-frequency turn-over power law & 13 / 85 (15.3\%) & 5 / 85 (5.9\%) \\
High-frequency cut-off power law & 5 / 85 (5.9\%) & 4 / 85 (4.7\%) \\
Log-parabolic spectrum & 3 / 85 (3.5\%) & 23 / 85 (27.1\%) \\
Double turn-over spectrum & 1 / 85 (1.2\%) & 1 / 85 (1.2\%) \\
Non-categorized & 1 / 85 (1.2\%) & 2 / 85 (2.4\%) \\
\hline\hline
    \end{tabular}
\end{table*}

\begin{table*}
    \caption{\justifying Bayes factor results for the 84 categorized MSPs. For each best-fit model, the table reports the number and percentage of pulsars with a decisive preference ($\ln\mathrm{BF} \ge 5$) over the second-best model and over the simple power law.}
    \label{tab:msp_dataset_betterby}
    \centering
    \begin{tabular}{ccc}
    \hline\hline
        Best-fit model & \makecell{$\ln\mathrm{BF} \ge 5$ \\ vs.\ second-best model}& \makecell{$\ln\mathrm{BF} \ge 5$ \\ vs.\ simple power law}\\
        \hline
Broken power law & 18 / 84 (21.4\%) & 28 / 84 (33.3\%) \\
Low-frequency turn-over power law & 1 / 84 (1.2\%) & 7 / 84 (8.3\%) \\
High-frequency cut-off power law & 1 / 84 (1.2\%) & 4 / 84 (4.8\%) \\
Log-parabolic spectrum & 1 / 84 (1.2\%) & 3 / 84 (3.6\%) \\
Double turn-over spectrum & 0 / 84 (0.0\%) & 1 / 84 (1.2\%) \\
Simple power law & 0 / 84 (0.0\%) & --- \\
        \hline\hline
    \end{tabular}
\end{table*}

\subsection{Spectra peaked at gigahertz frequencies}

GPS pulsars are characterized by a flux density maximum near 1~GHz. This behavior is commonly attributed to thermal free--free absorption associated with interaction between the pulsar and its surrounding environment \citep{Kijak_2007,rajwade_gigahertz_2016,Kijak_2017}.
Our systematic search for GPS pulsars, which we define as sources whose flux density peaks between 0.6 and 2.0 GHz, has significantly expanded the known population. By applying stringent criteria that require both the observed data and the best-fit model to show a clear peak within this range, we have identified 74 confident GPS candidates.

\begin{figure}
    \centering
    \includegraphics[width=\linewidth]{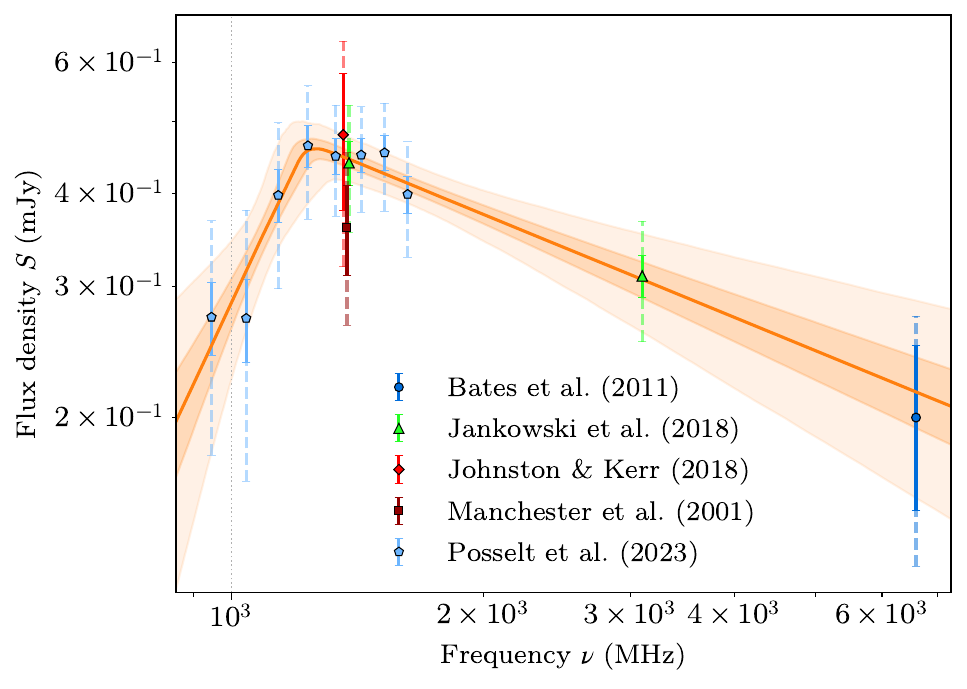}
    \caption{\justifying 
As \Fig{fig:J1809-1917}, but for the broken power law spectrum of PSR J1406--6121, one of the newly identified GPS pulsars in this work.
Both its observed data peak and the peak of the best-fit broken power law model lie near 1~GHz.
Solid error bars show the uncertainties reported in the literature, while dashed error bars indicate the rescaled uncertainties with inferred $e_{\text{fac}}$ parameters.
    }
    \label{fig:J1413-6141}
\end{figure}

A majority (58 out of 74) are new discoveries not listed in previous catalogues by \citet{kijak_low-frequency_2021} or \citet{xu_statistical_2024}. An example is PSR J1413$–$6141 (\Fig{fig:J1413-6141}), which shows a well-defined peak near 1 GHz in its data and broken power-law model.
Conversely, we find that 37 previously proposed GPS pulsars do not meet our dual-requirement criteria. This discrepancy likely arises from irregular data sampling or outlier measurements that can bias model fits when not rigorously vetted.

All 74 GPS pulsars, with both observed and modelled peak frequencies ($\nu_{\textrm{p}}$), are listed in \Table{tab:gps_pulsars}.
The dominant model for these sources is the broken power law, which best describes the spectra of 47 of the 74 GPS pulsars. This provides a clear, empirical basis for future studies aiming to understand the physical origin of these peaks, which is often attributed to processes like thermal free-free absorption in the pulsar's local environment.

\begin{table*}
    \caption{The list of 74 GPS pulsars identified in this work, with their peak frequency $\nu_{\rm p}$ and best-fit model.}
    \label{tab:gps_pulsars}
    \centering
\begin{threeparttable}
\makebox[\linewidth]{ %
\begin{tabular}{cccc|cccc}
\hline\hline
PSR name & Observed $\nu_{\rm p}$ & Model $\nu_{\rm p}$ & Model & PSR name & Observed $\nu_{\rm p}$ & Model $\nu_{\rm p}$ & Model \\
& (MHz) & (MHz) & & & (MHz) & (MHz) & \\
\hline
J0818--3232 & 1283.5 & 901.6 & double turn-over & J1701--4533 & 944.0 & 740.7 & low-freq turn-over PL \\
J0900--3144 & 944.6 & 805.4 & low-freq turn-over PL & J1705--3950*† & 1278.8 & 969.0 & low-freq turn-over PL \\
J0901--4624 & 944.0 & 849.4 & broken PL & J1718--3825* & 1038.0 & 1035.7 & low-freq turn-over PL \\
J0908--4913† & 945.0 & 953.2 & broken PL & J1721--3532 & 1435.0 & 1163.6 & broken PL \\
J1015--5719 & 1038.0 & 974.1 & broken PL & J1723--3659* & 944.0 & 781.1 & broken PL \\
J1019--5749 & 1523.0 & 1676.8 & low-freq turn-over PL & J1737--3555 & 728.0 & 781.3 & broken PL \\
J1020--6026 & 1382.0 & 1573.3 & low-freq turn-over PL & J1738--3211 & 728.0 & 964.2 & broken PL \\
J1048--5832 & 843.0 & 926.2 & low-freq turn-over PL & J1746--2856 & 1235.0 & 1231.8 & broken PL \\
J1055--6028 & 945.0 & 1274.2 & double turn-over & J1750--3157 & 606.0 & 712.9 & broken PL \\
J1110--5637 & 843.0 & 1591.1 & double turn-over & J1801--2304 & 1040.0 & 1299.0 & broken PL \\
J1112--6103 & 1360.0 & 1382.7 & low-freq turn-over PL & J1803--2137*† & 1315.0 & 857.2 & low-freq turn-over PL \\
J1126--6054 & 1232.0 & 1223.8 & broken PL & J1806--2125 & 1170.0 & 1119.2 & broken PL \\
J1138--6207 & 1360.0 & 1231.8 & broken PL & J1810--5338 & 660.0 & 1335.0 & broken PL \\
J1316--6232 & 1429.0 & 1685.2 & low-freq turn-over PL & J1815--1738 & 1360.0 & 1093.5 & broken PL \\
J1320--3512 & 1435.0 & 1626.0 & broken PL & J1818--1422† & 843.0 & 802.2 & low-freq turn-over PL \\
J1341--6220 & 1041.0 & 1222.8 & broken PL & J1820--1529 & 1278.5 & 1281.0 & broken PL \\
J1406--6121 & 1360.0 & 1232.0 & broken PL & J1822--1400* & 610.0 & 769.0 & broken PL \\
J1410--6132 & 1360.0 & 1383.4 & broken PL & J1826--1334*† & 1606.0 & 848.6 & broken PL \\
J1413--6141 & 1234.0 & 1201.9 & broken PL & J1828--1101 & 1332.0 & 1267.0 & low-freq turn-over PL \\
J1413--6222 & 1040.0 & 1106.8 & broken PL & J1832--1021† & 1435.0 & 961.1 & broken PL \\
J1452--6036 & 843.0 & 827.1 & broken PL & J1833--0827* & 843.0 & 782.7 & broken PL \\
J1513--5739 & 944.0 & 947.6 & broken PL & J1837--0604 & 1360.0 & 1534.8 & low-freq turn-over PL \\
J1513--5908† & 944.0 & 868.9 & broken PL & J1839--0321 & 1382.0 & 1213.5 & broken PL \\
J1514--5925 & 1332.0 & 1362.5 & low-freq turn-over PL & J1839--0643 & 1041.0 & 1020.9 & double turn-over \\
J1524--5706† & 945.0 & 965.0 & broken PL & J1841--0425 & 610.0 & 946.5 & broken PL \\
J1538--5551 & 1360.0 & 1035.3 & low-freq turn-over PL & J1841--0524 & 1041.0 & 1079.2 & broken PL \\
J1550--5418* & 1628.0 & 1695.1 & low-freq turn-over PL & J1843--0355 & 1332.0 & 1588.1 & low-freq turn-over PL \\
J1551--5310 & 1360.0 & 1188.1 & low-freq turn-over PL & J1844--0538† & 606.0 & 744.3 & broken PL \\
J1611--5209 & 944.0 & 906.1 & broken PL & J1847--0438 & 1429.0 & 1429.0 & broken PL \\
J1627--4706 & 1334.0 & 1257.1 & broken PL & J1850--0026 & 1041.0 & 1071.3 & broken PL \\
J1630--4733 & 1360.0 & 1264.8 & broken PL & J1852--0635*† & 1278.8 & 1279.3 & broken PL \\
J1632--4621 & 1038.0 & 1001.8 & broken PL & J1903+0327 & 1400.0 & 1121.0 & broken PL \\
J1632--4757 & 1360.0 & 1300.1 & broken PL & J1917+0834 & 943.0 & 733.6 & broken PL \\
J1636--4440 & 1360.0 & 1130.4 & low-freq turn-over PL & J2007+2722* & 1041.6 & 998.0 & low-freq turn-over PL \\
J1637--4642 & 1382.0 & 1115.7 & low-freq turn-over PL & J2124--3358 & 730.0 & 984.0 & broken PL \\
J1648--4611 & 1524.0 & 1155.2 & broken PL & J2234+0611 & 1400.0 & 866.7 & low-freq turn-over PL \\
J1653--4249 & 944.0 & 963.5 & broken PL & J2234+0944 & 944.6 & 653.3 & low-freq turn-over PL \\
\hline\hline
\end{tabular}}
\begin{tablenotes}
\item *: Pulsar previously identified as a GPS pulsar by \citet{kijak_low-frequency_2021}.
\item †: Pulsar previously identified as a GPS pulsar by \citet{xu_statistical_2024}.
\end{tablenotes}
\end{threeparttable}
\end{table*}

\begin{figure}
    \centering
\includegraphics[width=\linewidth]{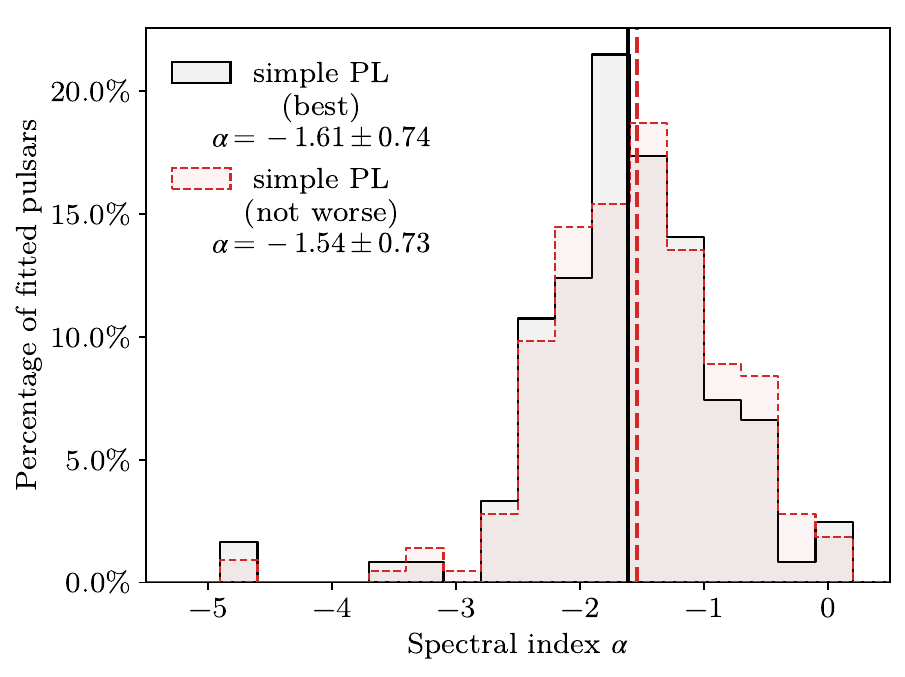}
    \caption{\justifying 
Distribution of mean spectral indices for 121 pulsars best fit by the simple power law, and for 214 pulsars whose simple power law fits are not evidently worse than other models (any $\ln \mathrm{BF} \le 3$).
    }
    \label{fig:alpha_distribution}
\end{figure}

\begin{figure}
    \centering
    \includegraphics[width=\linewidth]{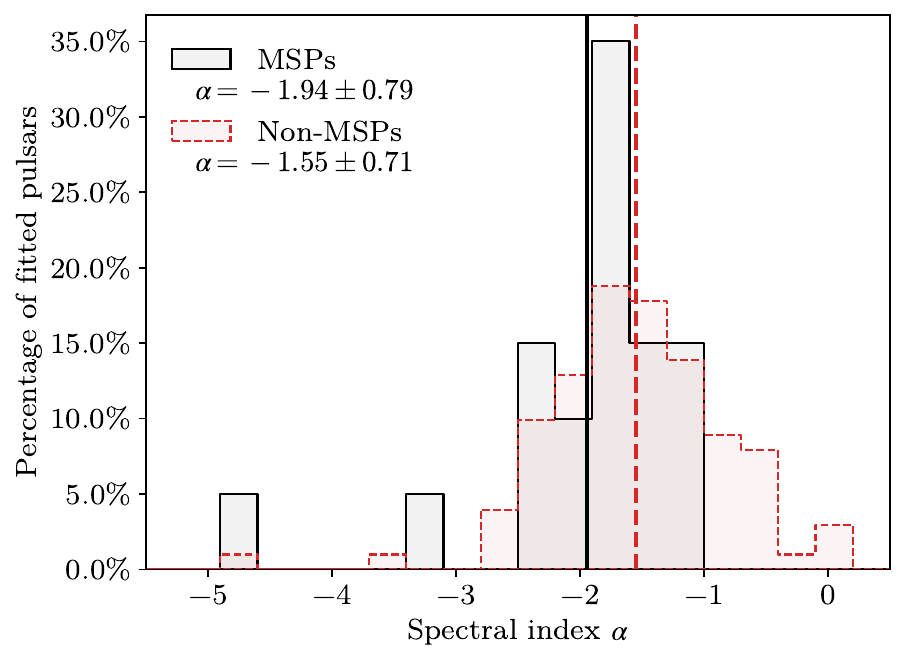}
    \caption{\justifying 
Distribution of mean spectral indices for 20 MSPs and 101 non-MSPs best fit by the simple power law. MSPs are steeper, but the difference is small ($\lesssim  1\sigma$).
    }
    \label{fig:alpha_msp}
\end{figure}

\begin{figure}
    \centering
    \includegraphics[width=\linewidth]{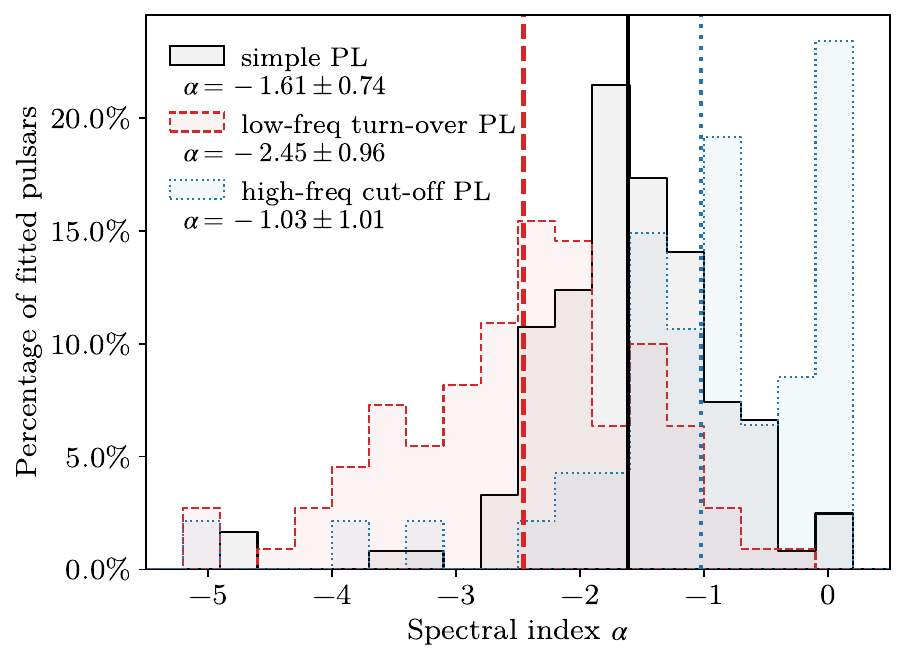}
    \caption{\justifying 
Distributions of mean spectral indices for three best-fit categories: simple power law (121), low-frequency turn-over power law (110 pulsars), and high-frequency cut-off power law (47 pulsars).
    }
    \label{fig:alpha_comparison}
\end{figure}

\subsection{Distribution of spectral indices}

Our analysis confirms and refines the established distribution of pulsar spectral indices. For the 121 pulsars best fit by a simple power law, the spectral indices follow a Gaussian distribution with a mean of $-1.61$ and a standard deviation of $0.74$ (\Fig{fig:alpha_distribution}), consistent with previous studies \citep[e.g.][]{Jankowski_2018,bates_pulsar_2013}.
Furthermore, our analysis confirms a modest trend toward steeper spectral indices for MSPs. For the 20 MSPs best fit by a simple power law, the mean spectral index is $-1.94$, compared with $-1.55$ for non-MSPs in the same category (see \Fig{fig:alpha_msp}).

Crucially, we find that the choice of spectral model significantly impacts the inferred spectral index. As shown in \Fig{fig:alpha_comparison}, the distributions for curved models are distinct.
Pulsars fit with a low-frequency turn-over model have a steeper mean index of $-2.45$ and a standard deviation of $0.96$.
Those fit with a high-frequency cut-off have a flatter mean index of $-1.03$ and a standard deviation of $1.01$.

This demonstrates that forcing a simple power-law fit onto a curved spectrum can systematically bias the measured spectral index. The steep indices associated with turn-over models may reflect the underlying spectrum before low-frequency absorption, while the flatter indices of cut-off models may indicate the onset of a high-frequency attenuation process. These distinct distributions provide new, model-dependent insights for linking spectral shapes to their physical origins.

\section{Summary}
\label{sec:conclude}

This work presents a comprehensive re-evaluation of pulsar radio spectra, leveraging the largest curated dataset of calibrated flux densities to date and a robust Bayesian framework. Our analysis fundamentally revises the prevailing understanding of pulsar emission characteristics.

Our principal conclusion is that the simple power law is not the dominant spectral form. Instead, complex spectra with curvature or breaks are the norm, with 68.8\% of pulsars decisively better fit by such models. This paradigm shift, starkly contrasting with the previous consensus established by \citet{Jankowski_2018}, is driven by our expanded dataset and the identification of a statistical bias in the previously used AIC metric, which artificially penalized complex models.

Key specific findings include:

\begin{enumerate}
    \item The broken power law emerges as the most common spectral shape, accounting for 60.1\% of pulsars in our Bayesian analysis.

\item Millisecond pulsars are not spectrally simple; more than half exhibit decisive curvature, primarily described by broken power laws, and they show a slightly steeper mean spectral index than normal pulsars.

\item We have identified 74 confident GPS pulsars, more than quadrupling the reliably known population, with the broken power law being the predominant model for these sources.

\item The distributions of spectral indices are model-dependent. The mean index for simple power laws is $-1.61$, but it steepens to $-2.45$ for low-frequency turn-overs and flattens to $-1.03$ for high-frequency cut-offs, indicating that forcing a power-law fit can introduce significant bias.
\end{enumerate}

In summary, this study demonstrates that pulsar spectra are far more complex and diverse than previously recognized. These findings challenge the community to move beyond the simple power-law approximation and provide a critical, model-classified foundation for future theoretical work aimed at understanding the underlying emission and absorption processes in pulsar magnetospheres and their environments.

\begin{acknowledgments}
We thank N. A. Swainston for his dedication in maintaining the pulsar flux density catalog.
We thank Marcus Lower, Andrew Zic, Zi-Yang Wang, Jingwang Diao for useful discussions.
This work is supported by the National Natural Science Foundation of China (Grant No.~12203004), the National Key Research and Development Program of China (No. 2023YFC2206704), the Fundamental Research Funds for the Central Universities, and the Supplemental Funds for Major Scientific Research Projects of Beijing Normal University (Zhuhai) under Project ZHPT2025001.
You, Z.-Q. is supported by the National Natural Science Foundation of China (Grant No.~12305059); The Startup Research Fund of Henan Academy of Sciences (No.~241841224); The Scientific and Technological Research Project of Henan Academy of Science (No.~20252345003); Joint Fund of Henan Province Science and Technology R\&D Program (No.~235200810111); Henan Province High-Level Talent Internationalization Cultivation (No.~2024032).
\end{acknowledgments}

\bibliography{refs}

\appendix

\section{Discussion on the AIC classification in \citet{Jankowski_2018}} \label{app:reproduction}

\citet{Jankowski_2018} published a dataset containing observations of 441 pulsars at three central frequencies: 728~MHz, 1382~MHz, and 3100~MHz. They also provided spectral classifications for these pulsars using an extended dataset that included additional measurements from the literature. In their analysis, the simple power law was reported as the best-fitting model for 62.6\% of the 441 observed pulsars and for 79.1\% of the 349 classified pulsars. The log-parabolic spectrum was the second most common model (35 pulsars). These results contributed to the widely adopted view that pulsar flux density spectra are generally well described by a simple power law, motivating extensive work on spectral index correlations based on this model.

To verify the classifications of \citet{Jankowski_2018}, we reconstructed their dataset following the procedures described in their paper. Our reproduction draws from the ATNF Pulsar Catalogue (version 1.54, the same version as used in Ref. \cite{Jankowski_2018}), flux density measurements reported in \citet{Jankowski_2018}, and the additional literature sources they cited. A notable limitation is that, although \citet{Jankowski_2018} list multiple measurements at individual frequencies, the dataset included with their publication averages these into a single value per frequency. As a result, our reproduction necessarily reflects the available published data rather than the full multi-channel dataset. ATNF data were retrieved using the Python package \textsc{psrqpy} \citep{pitkin_psrqpy_2018}, and other literature values were obtained via the \textsc{pulsar\_spectra} package \citep{swainston_pulsar_spectra_2022}. Missing entries, particularly from \citet{sieber_pulsar_1973} and \citet{maron_pulsar_2000}, were manually added from the original publications.

\begin{table*}
    \caption{\justifying
Classification of 441 pulsars according to table~5 of \citet{Jankowski_2018}, compared with results obtained using our reproduced dataset. Excluding unclassified pulsars, the simple power law accounts for 79.1\% of sources in \citet{Jankowski_2018} and 64.1\% in our reproduction.
    }
    \label{tab:jan_reproduction}
    \centering
    \begin{tabular}{ccc}
        \hline\hline
        Model & Classification in \citet{Jankowski_2018}  & Our reproduction   \\
        \hline
        Simple power law                  & 276 / 441 (62.6\%) & 202 / 441 (45.8\%)\\
        Broken power law                  & 25 / 441 (5.7\%)   & 44 / 441 (10.0\%)\\
        Log-parabolic spectrum            & 35 / 441 (7.9\%)   & 43 / 441 (9.8\%)\\
        High-frequency cut-off power law  & 3 / 441 (0.7\%)    & 14 / 441 (3.2\%)\\
        Low-frequency turn-over power law & 10 / 441 (2.3\%)   & 12 / 441 (2.7\%)\\
        (Unclassified)                    & 92 / 441 (20.9\%)  & 126 / 441 (28.6\%)\\
        \hline\hline
    \end{tabular}
\end{table*}

\begin{figure}
    \centering
    \includegraphics[width=\linewidth]{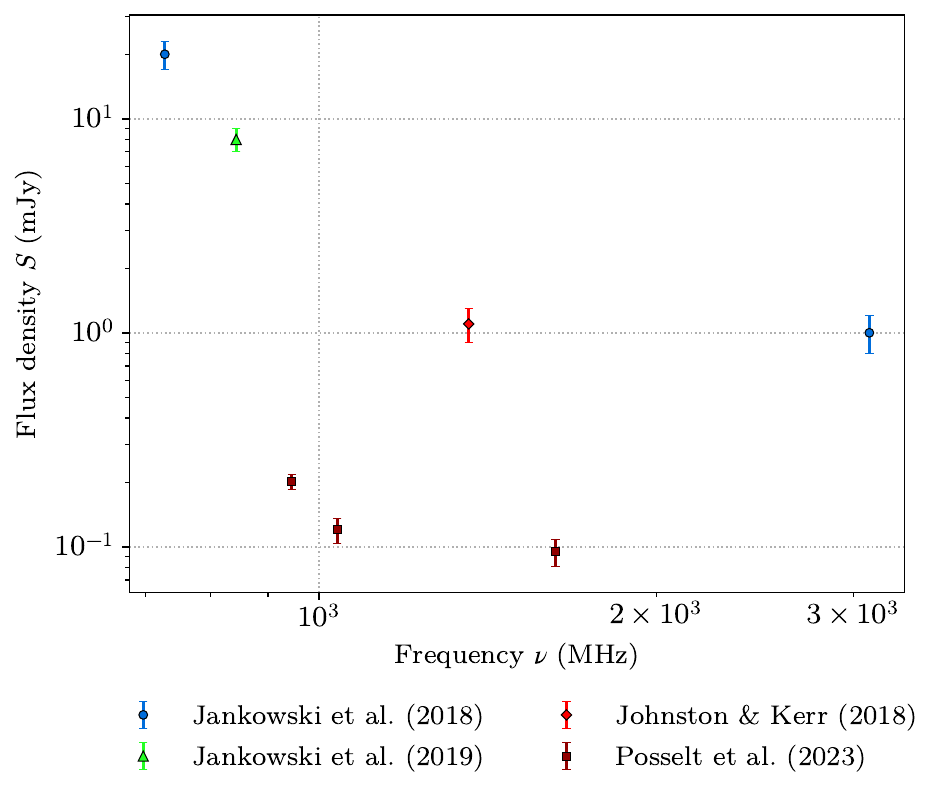}
    \caption{\justifying
    Spectrum of PSR~J1717$-$4054, which can be not described by any of the six models adopted in this work.}
    \label{fig:J1717-4054}
\end{figure}

\begin{figure}
    \centering
    \includegraphics[width=\linewidth]{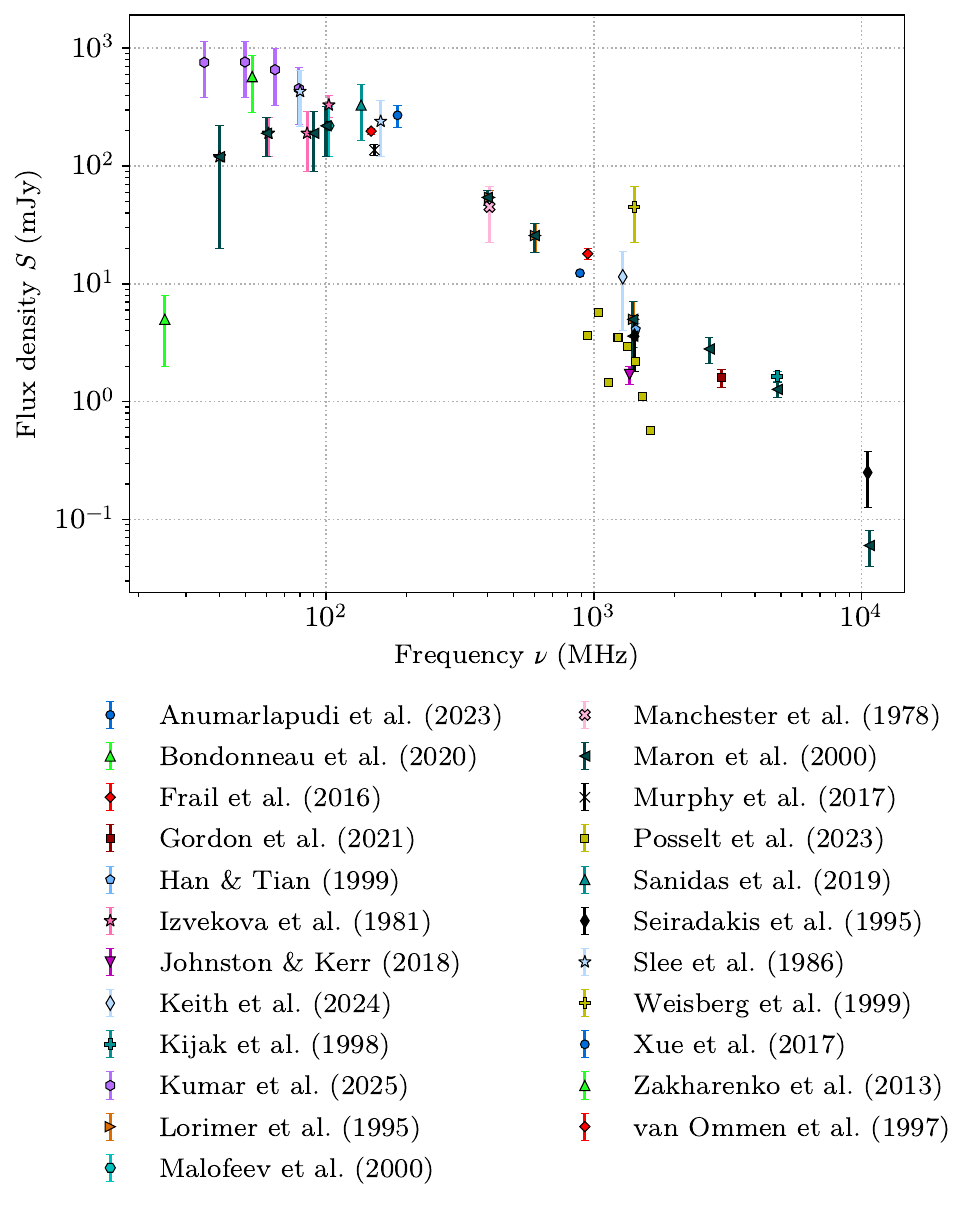}
    \caption{\justifying
    Spectrum of PSR~J1607$-$0032. Eight measurements with very small reported uncertainties from \citet{Posselt_2023} form a local turn-over feature that is inconsistent with the general trend.}
    \label{fig:J1607-0032}
\end{figure}

We applied the frequentist goodness-of-fit method of \citet{Jankowski_2018} to our reproduced dataset. The fitting procedure used maximum-likelihood and least-squares methods implemented via the \textsc{minuit} C++ library and its Python interface \textsc{iminuit} \citep{james_minuit_1975,dembinski_scikit_hepiminuit_2022}, together with the Huber loss function and the ``corrected AIC'' statistic introduced by \citet{Jankowski_2018} and implemented in \textsc{pulsar\_spectra}. We reproduced their fitting requirements as closely as possible. As shown in \Table{tab:jan_reproduction}, 202 of the 315 classified pulsars (64.1\%) prefer the simple power law, in good agreement with \citet{Jankowski_2018}. Although our dataset contains slightly fewer measurements because their multi-channel data were not available, the simple power law remains the most common best fit among the five models considered.

Despite validating the overall trend, we noted an intrinsic limitation of the corrected AIC:
\begin{equation}
    \mathrm{AIC}_c = -2 \ln L_{\max} + 2K + \frac{2K(K+1)}{N-K-1}, \label{eq:aic}
\end{equation}
where $L_{\max}$ is the maximum likelihood, $N$ is the number of measurements, and $K$ is the number of model parameters. The final term corrects for finite sample sizes. However, when $N-K-1$ equals zero, the correction term diverges, rendering the AIC infinite and disqualifying the model regardless of fit quality. For pulsars with only 5 measurements, all 4-parameter models (broken power law and low-frequency turn-over power law) are automatically excluded; with 4 measurements, even 3-parameter models are eliminated. This likely contributes to the strong preference for the simple power law in the original analysis and in our reproduction.

To mitigate this structural disadvantage against $(N-1)$-parameter models when fitting $N$-point spectra, we recalculated the AIC without the correction term. Under this adjustment, the fraction of pulsars classified as simple power laws drops to 26.8\%. In particular, 84 pulsars with only 4 or 5 measurements---previously forced into the simple power law category---shift to other models: 47 to the broken power law, 31 to the log-parabolic spectrum, 5 to the high-frequency cut-off power law, and 1 to the low-frequency turn-over model. This reclassification suggests that the dominance of the simple power law reported by \citet{Jankowski_2018} may be, at least in part, an artifact of their method penalizing multi-parameter models in sparsely sampled spectra.

\section{Not categorized pulsars}
\label{sec:bad-fits}

In \Table{tab:our_dataset_bestcat}, 13 pulsars cannot be assigned a spectral model under our Bayesian procedure described in \Sec{sec:bayesian_model_selection}. These cases can be divided into two groups.

Some pulsar spectra can not be described by any of the six models considered in this work. This may result from intrinsic flux variability, temporal evolution due to scattering or scintillation effects in the interstellar medium, or systematic errors arising from observational issues. Such cases include PSR~J0151$-$0635, PSR~J0943+1631, PSR~J1453+1902, PSR~J1652+2651, PSR~J1717$-$4054 (\Fig{fig:J1717-4054}), and PSR~J1747$-$2958.
    
For some pulsars, the Bayesian inference is largely dominated by a subset of measurements with very small errors. Increasing the prior upper bound of $e_{\text{fac}}$ may solve this issue. However, since only 7 pulsars are found in this group, including J0034$-$0721, J0152$-$1637, J0304+1932, J0630$-$2834, J1136+1551, J1607$-$0032 (\Fig{fig:J1607-0032}), and J1751$-$4657, we leave the investigation of these pulsars to a future work.

The frequentist AIC analysis identifies a partially different set of 14 pulsars that are not categorized.
Five pulsars ``fail" in \emph{both} Bayesian and frequentist analyses: J0151$-$0635, J0943+1631, J1136+1551, J1717$-$4054, and J1747$-$2958.
Among the remaining not-categorized AIC pulsars, J1246+2253 has apparently inconsistent flux densities, and for the others with a small number of measurements, the finite-sample AIC correction term diverges when $N - K - 1 = 0$.
In this latter case, models with fewer parameters are selected but they actually provide inadequate fits.
The Bayesian inference was successful for these pulsars because of the inclusion of the $e_{\text{fac}}$ parameter.
The full classification details of all 897 pulsars can be found in the \textsc{spectral\_fit} repository.

\end{document}